\documentclass[aps,prb]{revtex4}
\usepackage{graphicx}% Include figure files
\usepackage{bm}% bold math
\usepackage{nicefrac}
\usepackage{amsmath}
\newcommand{\be}{\begin{equation}}
\newcommand{\ee}{\end{equation}}
\newcommand{\bea}{\begin{eqnarray}}
\newcommand{\eea}{\end{eqnarray}}
\newcommand{\beb}{\begin{eqnarray*}}
\newcommand{\eeb}{\end{eqnarray*}}

\DeclareMathOperator*{\equivalent}{\sim}
\begin{document}
%\preprint{LPTMS-xxx}

\title{Edge properties of principal fractional quantum Hall states in the cylinder geometry}

\author{Paul Soul\'e}
\author{Thierry Jolicoeur}
\affiliation{Laboratoire de Physique Th\'eorique et Mod\`eles statistiques,
Universit\'e Paris-Sud, 91405 Orsay, France}

\date{July 13th, 2012}
%%%%%%%%%%%%%%%%%%%%%%%%%%%%%%%%%%%%%%%%%%%%%%%%%%%%%%%%%%%%%%%%%%%%%%%%%%%
\begin{abstract}
We study fractional quantum Hall states in the cylinder geometry with open boundaries.
We focus on principal fermionic $\nu=1/3$ and bosonic $\nu=1/2$ fractions in the case of hard-core interactions.
The gap behavior as a function of the cylinder radius is analyzed.
By adding enough orbitals to allow for edge modes we show that it is possible
to measure the Luttinger parameter of the non-chiral liquid formed by the combination
of the two counterpropagating edges when we add a small confining potential.
While we measure a Luttinger exponent consistent with the chiral Luttinger theory prediction
for the full hard-core interaction, the exponent remains non-trivial in the Tao-Thouless limit
as well as for simple truncated states that can be constructed on the cylinder.
If the radius of the cylinder is taken to infinity the problem becomes a Tonks-Girardeau
one-dimensional interacting gas in Fermi and Bose cases.
Finally we show that the the Tao-Thouless and truncated states 
 have an edge electron propagator which decays spatially with a Fermi-liquid exponent
even if the energy spectrum can still be described by a non-trivial Luttinger parameter.
\end{abstract}
%\pacs{71.10.Pm,73.43.Cd}
\maketitle
%%%%%%%%%%%%%%%%%%%%%%%%%%%%%%%%%%%%%%%%%%%%%%%%%%%%%%%%%%%%%%%%%%%%%%%%%%%%

\section{introduction}

The physics of edge states~\cite{AHM90,Wen90,Wen92,Wen95}
 is very important in the description of the fractional quantum Hall effect
(FQHE).  Under a strong enough magnetic field, two-dimensional electron systems have various
incompressible liquid ground states with a gap to bulk excitations. However there are gapless modes 
on the boundary of the sample. These modes are crucial to the existence of quantized conductance
at the FQHE plateaus. Many experimental probes can be used to study this edge physics.
For example,
it is possible to induce tunneling from a Fermi liquid into a FQHE state or to  create tunnel coupling between
spatially separated states. Some evidence for the fractional charge of the basic elementary excitations
comes from noise measurement involving edge mode manipulations through a quantum point contact.

The edge states are expected to form a very special one-dimensional strongly interacting electronic system,
a chiral Luttinger liquid~\cite{AHM90,Wen90}. This picture has been studied in detail for the simplest $\nu =1/m$ FQHE states.
The necessary appearance of gapless edge modes has been linked to the existence of some kind of topological order present
in the bulk liquid state while there is no usual Landau-Ginzburg type of ordering. 
From the theoretical side,
the description has been based on the use of the Laughlin wavefunction which is an explicit first-quantized description
of the electronic ground state. The use of a powerful plasma analogy approach allows the computation of the
electron propagator along the edge of a FQHE droplet~\cite{Wen95}. In the case of the $\nu=1/3$ state, Wen has shown that the 
spatial decay
of this propagator involves a non-Fermi liquid exponent $\alpha =3$.  Extension of this approach
to other FQHE states of the prominent series of fractions $\nu =n /(2n+1)$ predict a constant exponent $\alpha =3$
independent of the filling factor.
This value is also obtained using composite fermion (CF) wavefunctions~\cite{Jolad07,Jolad2010a,Jolad2010b,Sreejith2011}. 
Experimentally~\cite{Chang03} it is possible to measure the I-V characteristics for tunneling from a Fermi liquid state
to a FQHE liquid and access the value of $\alpha$.
There is some evidence for exponent values that varies smoothly with the filling factor and which are not given by
the universal value $\alpha =3$. It is not clear yet if these deviations really represent a problem for the chiral Luttinger picture of the 
edge states. Direct numerical measurements are hampered by the small sizes of systems
that can be studied by exact diagonalization~\cite{Goldman01,Wan05}. 
In the calculation of the electron propagator there are essentially two key ingredients.
The first is the effective long-wavelength low-energy theory. This is the chiral single boson free theory
for principal Laughlin~\cite{Laughlin83} fractions $\nu=1/m$. The second ingredient is the expression of the electron
operator in terms of the bosonic modes. Doubts have been raised about the universal characteristics
of the electron operator given by Wen~\cite{Palacios96,Zulicke03}. Since the question has to do with long-distance properties
it is important to have results for model wavefunctions for a very large number of particles.
In addition to these basic potential problems, it is also known that the Coulomb interaction may lead
to edge reconstruction~\cite{Chang03} favoring a complex non-universal physics.

In this paper, we study the FQHE in the cylinder geometry by using exact diagonalization as well as some analytic
wavefunctions and discuss implications for the edge theory. The convenient way to formulate a FQHE problem
in this geometry is to impose a sharp cut-off in momentum space. This leads to a problem which is in one-to-one
correspondence with the sphere geometry but without the full rotation symmetry. We consider spinless fermions
at $\nu=1/3$ and bosons at $\nu=1/2$ that are the simplest examples of Laughlin FQHE fluids. The interactions
are taken to be hard-core.
If the number of available orbitals is chosen by taking into account the shift quantum number as in the spherical
geometry, we show that it is possible to study the physics of incompressible states as a function of the cylinder radius.
But  it has the advantage that
it is natural to think about the two boundaries of the system as supporting edge excitations provided 
 we add extra orbitals.
In the large radius limit at fixed number of particles we discuss the crossover to one-dimensional gapless physics.

In the Landau gauge it is convenient to add a parabolic potential along the axis of the
cylinder. This lifts the edge mode degeneracies and by combination of the two edges with opposite chiralities
we observe the formulation of a non-chiral Luttinger liquid. Since this liquid has now a small kinetic energy
it is possible to measure its Luttinger parameter $g$ by computing the momentum dependence of low-lying states.
As is expected from Wen's theory we find that $g=1/m$ in the case of the hard-core interactions when the system
has an ``aspect ratio'' suited to the thermodynamic limit. We also compute the Luttinger exponent for the simple
Tao-Thouless (TT) state which is also given by Wen's value. In the large radius limit there is a change in the Luttinger exponent
that we observe and explain in the Bose case. Finally we compute the edge electron propagator for the TT
state and for a special truncated state introduced recently~\cite{Soule2012} and show that it is given
by its Fermi-liquid value $\alpha=1$ and not the expected $\alpha=3$ even if these states have the correct
edge energetics as described by the chiral Luttinger liquid Hamiltonian.

In section \ref{geo}, we present the cylinder geometry and compare it with the well-studied spherical, planar
and toroidal geometries. 
The behavior of the gap in the cylinder geometry is discussed in section \ref{gapsection}.
In section \ref{edgecount}, the edge counting of various model states is studied.
In section \ref{Lut}, we add a parabolic potential well to lift the degeneracy of the edge states and compute
the Luttinger parameter $g$. 
The limit of very large cylinder radius is discussed in section \ref{hoop}.
The density profiles of model states  as well as correlations and
the electron propagator at the edge are studied in section \ref{edgecorr}. 
Finally our conclusions are presented in section \ref{conclude}.

%****************************************************************************************
\section{Geometries}
\label{geo}

Our study is restricted to FQHE states of spinless fermions and bosons residing in the lowest Landau level (LLL).
In the real world, the incompressible liquids are created by the interactions between particles
that are the Coulomb interaction for electrons and the pure s-wave low-energy scattering between
ultracold atoms for bosons~\cite{Chang05}. If we project an arbitrary two-body interaction into the LLL, then the Hamiltonian 
may be written as a sum of projection operators~:
\be
{\mathcal H}=\sum_{i<j} \sum_m V_m {\hat P}_{ij}^{(m)} ,
\label{pseudos}
\ee
where $m$ is a positive integer, the coefficients $V_m$ are the so-called Haldane pseudopotentials
and ${\hat P}_{ij}^{(m)}$ projects the pair $(ij)$ onto relative angular momentum $m$.
Quantum statistics leads to the fact that only odd (resp. even) $m$ values matter for fermions (resp. bosons).
This formula is valid for geometries in which one can define a relative angular momentum
like the unbounded plane or the sphere.
Ultracold atoms are described solely by $V_0$ while the Coulomb interactions have all nonzero
pseudopotentials. In this last case the $V_m$ are all positive and decrease monotonically  as a function of $m$.
It is known that the FQHE physics appears when there are dominant hard-core interactions and in fact
 the fermionic FQHE has many features in common with the pure hard core interaction with only
nonzero and positive $V_1$. In this work we will only consider such ultra-short range models.
These hard-core interactions can be defined for all geometries through the matrix elements of a model
two-body potential in real space before projection onto the LLL.

The LLL physics may be studied in different geometries and the gauge should be
chosen accordingly. One possibility is to use the symmetric gauge in the unbounded plane $A=\nicefrac{1}{2} B\times r$.
With the complex coordinates $z=x+iy$ the basis states can be taken as~:
\be
\phi_m(z)= \frac{1}{\sqrt{2^{m+1}\pi m!}}\,\,  z^m\, {\mathrm{e}}^{-|z|^2/4\ell^2},
\ee
where the magnetic length $\ell=\sqrt{\hbar c/eB}$ sets the scale of the problem. 
From now on, we will set $\ell =1$. In this geometry
an arbitrary many-body wavefunction in the LLL is given by a polynomial $P$ in the $z_i$ variables
times a universal exponential~:
\be
\Psi (z_1,\dots , z_N)= P(z_1,\dots ,z_N)\, {\mathrm{e}}^{-\sum_i|z_i|^2/4}.
\ee
The original Laughlin wavefunction is given by~:
\be
P=\prod_{i<j}(z_i-z_j)^3\, ,
\label{Laughlin}
\ee
where we have omitted the universal Gaussian factor.
It describes the physics of electrons at filling factor $\nu =1/3$.
This homogeneous polynomial is a zero energy ground state of the pure hard core $V_1$ model
and it is unique if we require the polynomial to be of smallest total degree. All the other zero energy eigenstates
are obtained by multiplying Eq.(\ref{Laughlin}) by a symmetric polynomial. This leads
to edge mode excitations for the Laughlin droplet and,
for degree of order $N$, to quasihole states. The Bose case is similar with now a power $m=2$ 
in the Jastrow factor in Eq.(\ref{Laughlin})
and for $\nu =1/2$ the $V_0$ model has the Laughlin wavefunction as the densest ground state.
These hard-core models have gapless quasiholes and their wavefunctions are known exactly
however this does not extend to quasielectron states obtained by removal of a flux quantum from the
ground state flux. Here only approximate albeit very precise wavefunctions are known
from composite fermion theory~\cite{JainBook}.

We now turn to the cylinder geometry where
we use the Landau gauge
with $A_x=0$ and $A_y=B x$.
Eigenstates can be chosen with definite momentum along the $y$-axis. 
If we 
impose periodic boundary conditions along the $y$ direction with a finite extent $L$~:
$\psi(y+L)\equiv \psi(y) $ the momentum $k$ is then quantized~: $k=2\pi n/L$
where $n$ is an integer. This defines the cylinder geometry, the radius of the cylinder being $L/2\pi$.
The cylinder geometry has been repeatedly mentioned in the FQHE literature. The 
Luttinger liquid theory has been studied in a field-theoretic way by Wen~\cite{Wen90}.
Chamon and Wen~\cite{Chamon94} have also used the cylinder geometry to study edge reconstruction
in the Hartree-Fock approximation. Milovanovic and Read have given an analytical description of cylinder edge
states of various Abelian and non-Abelian states~\cite{MR96}. The Laughlin wavefunction was studied
by Rezayi and Haldane~\cite{RH94}.
 The LLL one-body wavefunctions are given by~:
\begin{equation}
\phi_{n} (x,y)= 
\frac{1}{\sqrt{L {\ell} \sqrt{\pi}}}
{\mathrm{e}}^{-\frac{2\pi^2}{L^2}{ n^2}}
Z^n \,\,
{\mathrm e}^{\displaystyle{- {x^2}/{2 }}},
\quad
 Z\equiv {\mathrm e}^{\frac{2\pi}{L}(x + i y) }\, .
\end{equation} 
It is important to note that the power $n$ of the complex $Z$ coordinates can be positive or negative.
A generic many-body wavefunction is thus a polynomial in the $Z$s \textit{and} $Z^{-1}$s of the particles~:
\be
\Psi (Z_1,\dots,Z_N)={\mathcal P}(Z_1,\dots,Z_N) 
\prod_i {\mathrm e}^{\displaystyle{- {x_i^2}/{2 }}} .
\ee
The Laughlin wavefunction in the cylinder geometry has been written by Rezayi and Haldane~\cite{RH94}~:
\be
\Psi^{(m)} =\prod_{i<j} \left(\frac{Z_i^{\frac{1}{2}}}
{Z_j^{\frac{1}{2}}}-\frac{Z_j^{\frac{1}{2}}}{Z_i^{\frac{1}{2}}}\right)^m \, 
=\prod_{i<j}(Z_i-Z_j)^m \times \left(\prod_i Z_i\right)^{-m(N-1)/2}\, ,
\label{LPsi}
\ee
where we omit the ubiquitous exponential factor. The filling factor is then $\nu =1/m$
with $m$ odd (resp. even) for fermions (resp. bosons).

The cylinder geometry is invariant by global translations along the $x$-direction. As a consequence the Laughlin
droplet modeled by Eq.(\ref{LPsi}) can be translated by multiplication by the operator ${\hat U}^{p}$ where~:
\be
{\hat U}=\prod_i Z_i \, \quad 
\label{Uop}
\ee
 and $p=m(N-1)/2$ so that it takes the form familiar from the planar geometry with now
the simple substitution $z_i \rightarrow Z_i$. The shift of all occupied states by the operator $\hat U$ exactly
corresponds to the flux insertion in the planar case when one uses a solenoid at the origin of the coordinates.
Incompressible FQHE states are realized for a special matching of the number of particles and the number of orbitals
which involves the so-called {shift} quantum number. 
By using the shift operator ${\hat U}$ it is clear that in fact the apparition of negative powers of $Z$ variables
is unessential~: states can always be translated so that their minimum power is zero.
The Hilbert space is truncated by imposing $|n|\leq N_{max}$. Since
the Gaussian factor implies that there is spatial localization of orbitals, the system has ``quasi'' hard walls
at  $|x|=2\pi N_{max}/L$ and there are $2N_{max}+1$ orbitals. 
To consider states with an even number of orbitals it is necessary to slightly enlarge the definition
of wavefunctions and allow for antiperiodic one-body boundary conditions. This leads now to half-integer
$n$ in the allowed momenta.
This set of boundary conditions breaks explicitly the translation symmetry along the $x$-direction.
It also creates two physical boundaries that can support edge modes of FQHE states.
If we reason at fixed number of particles then there is a set of preferred densities when the system form
an incompressible liquid. They will appear as a unique ground state with a gap to all excitations for a definite
relationship between $2N_{max}$ and $N$. In this case edge excitations are blocked by the hard walls.
If we add extra orbitals then there will be low-lying modes including edge excitations as well as the global
translations of the ground state droplet. 
In section (\ref{edgecount}) we discuss the identification of these various modes.

Writing the interactions in the
 second-quantized language we have in the fermionic case~:
\begin{equation}
 \mathcal H_F = \frac{g_f\sqrt{2\pi^3}}{L^3}\sum_{\{n_i\}}  [(n_1-n_3)^2-(n_1-n_4)^2]\quad
\lambda^{ (n_1 - n_3)^2 + (n_1 - n_4)^2 }\quad
c^\dag_{n_1} c^\dag_{n_2} c_{n_3} c_{n_4} \, ,
\label{HamF}
\end{equation} 
where the interaction is $V(r)=g_f\Delta\delta^2(r)$
while for bosons with interactions $V(r)=g_b\delta^2(r) $ we find~:
\begin{equation}
 \mathcal H_B = 
\frac{g_b}{L\sqrt{8\pi}}
\sum_{\{n_i\}}  \quad
\lambda^{ (n_1 - n_3)^2 + (n_1 - n_4)^2 }\quad
b^\dag_{n_1} b^\dag_{n_2} b_{n_3} b_{n_4} \, ,
\label{HamB}
\end{equation} 
where
the sum is restricted to $n_1+n_2=n_3+n_4$. Creation and annihilation operators are denoted by $c_n, c_n^\dag$
for fermions and $b_n, b_n^\dag$ for bosons.
We have  defined the convenient parameter
$\lambda ={\mathrm{e}}^{-\frac{2\pi^2}{L^2}}$.
Many-body eigenstates of this problem can be classified according to their total momentum
$K$ along the $y$-direction. From now on, we measure the momentum in units of $2\pi/L$.
Note that the two-dimensional problem looks now like a one-dimensional chain of particles hopping
on sites indexed by the momentum $n$. This is due to the fact that the guiding center coordinates are
quantum-mechanically conjugate in the LLL. The momentum conservation of the original problem now
means center of mass position conservation if we think in terms of hopping of particles along the chain.
It is important to note that there is an explicit $L$-dependent factor in front of each of these hard-core Hamiltonians.

As a function of the periodic length $L$ there are two interesting limiting cases.
When $L\rightarrow 0$ we have the so-called Tao-Thouless~\cite{Tao83,Thouless84} or thin-torus limit. The orbitals are spatially separated
and their overlaps are decreasing exponentially fast. This leads to a hierarchy of interactions  and the leading contributions
are electrostatic-like repulsion. The system crosses over to a charge-density wave physics.
The ground state becomes a simple Slater determinant (in the Fermi case) with one fully
occupied state every three orbitals for the filling fraction $\nu =1/3$. This 
Slater determinant is in fact a Vandermonde determinant and can be computed in closed form.
The corresponding wavefunction is the Tao-Thouless
wavefunction~:
\be
\Psi_{TT}= \prod_{i<j}(Z_i^3-Z_j^3) \, .
\label{TTwavef}
\ee
It was proposed originally as a candidate wavefunction to describe the FQHE state at $\nu=1/3$.
However it has clearly  correlations that are very different from those of the Laughlin wavefunction.
It is obvious that the binding of the zeros to the particle locations does not exist in this TT state
and this binding is at heart of the topological order hidden in the FQHE state.
We will demonstrate in section (\ref{Lut}) that the edge of this state is a Fermi liquid unlike the true FQHE state.

In the opposite limit
when $L\rightarrow \infty$ at fixed number of particles, the liquid is now squeezed along a hoop
and the two edges are in close contact. This should be a regime describable by one-dimensional physics
arguments. 
The Laughlin wavefunction Eq.(\ref{Laughlin}) reduces in this limit to the Calogero wavefunction which is the ground state
of a well-studied integrable one-dimensional problem.
In between, when the cylinder has approximately a ``square'' shape with an aspect ratio
close to unity we expect that the interacting particles display the FQHE phenomenon.
There is numerical evidence from exact diagonalization that it is so~\cite{Soule2012}.
Extrapolation to large number of particles should lead to results pertaining to the thermodynamic limit
of the FQHE provided one finds independence upon the aspect ratio. This is not a priori guaranteed
and has plagued numerical studies on the torus i.e. with completely periodic boundary conditions.
The extent of the droplet along the $x$-direction is given by $\approx 2\pi N/(\nu\times L)$
and thus the aspect ratio is $\approx 2\pi N /(\nu L^2)$ so the proper scaling to access the bulk FQHE
is $L\approx N^{1/2}$. In the thermodynamic limit the incompressibility is related to the existence
of a non-zero gap in units of microscopic couplings $g_f$ or $g_b$ or the corresponding pseudopotentials.

We now discuss the expansion of interactions Eqs.(\ref{HamF},\ref{HamB})
 in powers of $\lambda$~:
\be
\mathcal H =\sum_r \lambda^r \mathcal H_r .
\ee
In the TT limit $\lambda\rightarrow 0$ the physics is dominated by the first terms of this expansion~\cite{Chui86}.
If we neglect all hoppings and just consider the classical problem with extended repulsions then it belongs to the
family of convex interactions as a function of the range that were studied by Hubbard~\cite{Hubbard78} 
and Pokrovsky and Uimin~\cite{PU78}. It is known that there is a rich structure of ground states as a function
of the filling factor with a devil staircase reminiscent of the hierarchy of quantum Hall fractions~\cite{BK,BHHK}. 
However the physics is quite different from that of the ordinary FQHE. In the classical
approximation the ground state is then a Slater determinant built from the classical minimum energy configuration 
(a permanent for bosonic states) which does not contain the Jastrow factor repulsion which is present in the
Laughlin wavefunction or the CF states
However the true FQHE problem, even for pure
hard-core interactions deviates from purely electrostatic models already at low order in $\lambda$ due to the appearance
of hopping terms with conserved center-of-mass~\cite{Seidel05}. It is thus natural to focus on  the truncated Hamiltonians that consistently
include the first nontrivial hopping terms. In the fermionic case we thus define~:
\begin{equation}
{\mathcal H}^{FT}=\lambda\sum_i {\hat n}_i {\hat n}_{i+1}
+4\lambda^4\sum_i {\hat n}_i {\hat n}_{i+2}
+9\lambda^9\sum_i {\hat n}_i {\hat n}_{i+3}
-3\lambda^5\left[\sum_i c_i^\dag c_{i+1}c_{i+2}c_{i+3}^\dag+h.c.\right].
\label{TF}
\end{equation}
The corresponding Bose Hamiltonian is given by~:
\begin{equation}
{\mathcal H}^{BT}=\sum_i {\hat n}_i({\hat n}_i - 1)
+4\lambda\sum_i {\hat n}_i {\hat n}_{i+1}
+4\lambda^4\sum_i {\hat n}_i {\hat n}_{i+2}
+2\lambda^2\left[\sum_i b_i^\dag b^2_{i+1}b_{i+2}^\dag+h.c.\right].
\label{TB}
\end{equation}

In these equations the momenta are written as a site index $i$,
$b_i,b_i^\dag$ (resp. $c_i,c_i^\dag $) are bosonic (resp. fermionic) operators and ${\hat n}_i$ is the occupation number. 
Here we focus on exact eigenstates that can be obtained analytically in closed form.
For simplicity we have removed the overall energy scale factors that appear in the problem
in Eq.(\ref{HamF},\ref{HamB}) but we will show that they are important later in this section.

The truncated Hamiltonians have infinitely many exact eigenstates given by simple formulas
in second-quantized language. They are of the form~:
\be
\Psi_{exact}=  {\mathcal S} | R \rangle ,
\label{exact}
\ee
where the operator $ {\mathcal S} $ performs a squeezing operation on the Fock states. Its precise
form depends upon the statistics of the particles~:
\begin{equation}
{\mathcal S}_F =\prod_n (1+3\lambda^4 c_{n-1}c_n^{\dag} c_{n+1}^\dag c_{n+2}),
\quad
{\mathcal S}_B =\prod_n (1-\lambda^2 b_{n-1}(b_n^{\dag})^2 b_{n+1}).
\label{SQ}
\end{equation}
The special state $|R\rangle$ is a ``root'' configuration i.e. a Slater determinant in the Fermi case
where particles have large enough spatial separation. For example if we take~:
\be
|R_F\rangle = |  1001001\dots 1001\rangle \, ,
\ee
we obtain the Fermi exact ground state at $\nu=1/3$ and similarly~:
\be
|R_B\rangle = |  1010\dots 0101\rangle \, ,
\ee
gives the Bose ground state at $\nu=1/2$. A similar construction can be performed
on the torus geometry~\cite{Nakamura}.

To accommodate a unique ground state root configuration in a finite number of orbitals
requires precisely the flux vs number of particles that includes a nontrivial shift~:
$2N_{max}=m(N-1)$ at filling $\nu=1/m$.
If we add more orbitals then we can construct a family of states by adding extra zeros at the boundaries of the system~:
\be
\Psi = {\mathcal S}|00\dots 00\rangle \otimes |R_{F,B}\rangle \otimes |0\dots \rangle
\ee
This is at no energy cost and is a reflection of the center of mass degeneracy in the LLL.
The global shift operator $\hat U$ precisely translates the ground state pattern $|R_{F,B}\rangle$ in the available
``free'' space given by the extra orbitals. We can also insert extra zeros inside the bulk pattern defining the ground state.
These states are still exact zero-energy states and they are quasihole excitations. In section \ref{edgecount} we 
show that the  number of edge states from this construction is in agreement with edge theory.
The special form of the truncated states is similar to valence bond states and this may be used
to investigate the influence of disorder~\cite{Monthus98} which is difficult by exact diagonalization techniques.

In the TT limit we have $\lambda\rightarrow0$ and the wavefunctions obtained through
formulas Eq.(\ref{SQ}) become very simple~: in the Fermi case they reduce to
  Slater determinant  and to permanents in the Bose case.
In the general case these special states when expanded in the occupation number Fock basis
span only a small fraction of the total available configuration space. There is a ``root'' configuration
from which all state can be generated by acting with squeezing operations~\cite{BH08}.
This structure is also shared by the Laughlin wavefunction. However this latter state involves
squeezing at arbitrary distances while our truncated states are generated through nearest-neighbor squeezes.
As we show in section \ref{edgecorr}, they lack the nontrivial edge correlations due to the topological order
built in the Laughlin wavefunction.

\section{Gap behavior}
\label{gapsection}

%%%%%%%%%%%%%%%%%%%%%%%%%%%%%%%%%%%%%%%%%%%%%%%%%%%%%%%%%%

\begin{figure}[htb]
\includegraphics[width=0.8\columnwidth]{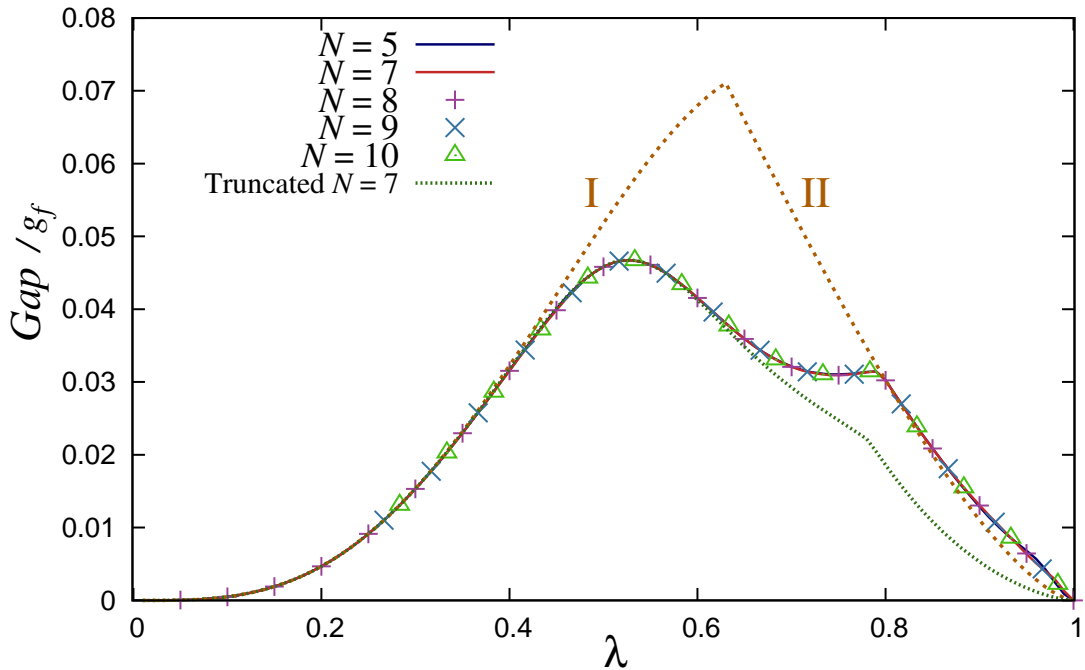}
\caption{The gap in units of $g_f$ as a function of the cylinder diameter parameterized by
the coupling $\lambda=\exp (-2\pi^2/L^2)$. The fraction $\nu=1/3$ is realized by taking exactly
$3(N-1)+1$ orbitals and the gap is computed between the $K=0$ ground state and the
lowest lying excited state regardless of its momentum. The values for the hard-core interaction are given by the
continuous line for N=5,7 fermions and by symbols for N=8,9,10. The truncated Hamiltonian has a similar behavior
as displayed by the dotted line for N=7.
There is a plateau near $\lambda\approx 0.6-0.8$ with a value close to what we expect be the bulk 
gap at $\nu=1/3$ for the hard-core model.
Curve I is the gap value in the TT limit while curve II is the gap of the magnetoroton for the truncated model.
The truncated model has a gap which is always lower than curve II because of $K=0$ lower-lying states.}
\label{Gap}
\end{figure}

%%%%%%%%%%%%%%%%%%%%%%%%%%%%%%%%%%%%%%%%%%%%%%%%%%%%%%%%%%

It is well known that the principal FQHE states have a nonzero gap in the thermodynamic limit which
is governed by the so-called magnetoroton branch of low-lying excited states. These states can be observed
without changing the flux. In the CF language they can be described as the excitation of one composite fermion
from the CF LLL to the first excited CF Landau level. In the spherical geometry such excitations can be classified
by total angular momentum $L_{tot}=N,N-1,N-2,\dots$ till the magnetoroton branch merges with higher excited state
which happens for small non-universal values of $L_{tot}$. In the cylinder geometry we expect that this branch
should be prominent and that its degeneracy is lifted since there is no complete rotation symmetry. The momentum conservation
is equivalent to $L_z^{tot}$ conservation on the sphere.  The splitting of the rotation multiplet leads
thus to values of $K=N,N-1,\dots N-k$ with $k+1$ almost degenerate states. This is what we observe in the
case of the hard-core interaction. In the truncated model, while this quasi-degeneracy is still present some
of the members of the branch are given by exact formulas. The magnetoroton with maximal momentum  $K$
value is given by~:
\begin{equation}
\Psi_{MR} = \mathcal{S}_F|11000100100\dots 0010\rangle \, .
\label{MR}
\end{equation}
It has a gap equal to $\lambda$ in the units of the truncated Hamiltonian Eq.(\ref{TF}) and can be thought as a 
quasiparticle-quasihole pair where the quasiparticle is the excess density at one end of the cylinder and the quasihole is
the deficit at the other end. By changing the location of the extra zero in the bulk pattern we generate a set of states
with $K=N,N-1,\dots N-k$ that share exactly the same gap.

We now discuss the behavior of the gap as a function of the length of the cylinder. We use a number of orbitals
that leave no space for edge excitations i.e. $2N_{max}=m(N-1)$ for principal Laughlin fluids. Fig. (\ref{Gap})
shows the value of the gap between the isolated $K=0$ ground state and the first excited state at the same flux
regardless of the momentum of this excited state. For aspect ratios in the range $\lambda\approx 0.6-0.8$ we observe a plateau behavior
that should extrapolate smoothly to the thermodynamic limit provided we send $N\rightarrow\infty$ and 
$L\rightarrow\infty$ at fixed $N/L^2$. Our limited data for the hard-core model are in fact quite insensitive to $N$ in this range
of $\lambda$. The value of the plateau is  $\approx 0.03g_f$ close to the present estimate
of the bulk gap for the hard-core model $\approx 0.033g_f = 0.41 V_1$.

First we note that there is an interesting limiting behavior in the TT limit $\lambda\rightarrow0$
where we find that the ground state continuously evolves to the Slater state $|1001001\dots\rangle$
and the excited states are governed by the second-nearest-neighbor interaction in Eq.(\ref{HamF}).
Taking into account the dimensional factors in the second-quantized Hamiltonian it means that this energy is
given by  $16g_f\sqrt{2\pi^3}\lambda^4/L^3$ leading to an asymptotic behavior
$\lambda^4|\log\lambda|^{3/2}$ as $\lambda\rightarrow 0$. This is shown as curve I in Fig.(\ref{Gap}).

When $L\rightarrow\infty$ i.e. $\lambda\rightarrow1$ this is the ``hoop'' limit where the system is reduced to bare edges
and we expect to find one-dimensional physics with gapless excited states. This is very easy to establish in the truncated model.
Indeed the composite fermion branch with root configuration $|11000100100\dots\rangle$
that corresponds to the magnetoroton excitations has an energy which is given by
$4g_f \sqrt{2\pi^3}\lambda/L^3$ in physical units. Hence this subset of the magnetoroton branch goes to zero energy in the hoop limit
as $|\log\lambda|^{3/2}$.  This is curve II in Fig.(\ref{Gap}).
This means that the truncated model is gapless if we take the thermodynamic limit sending the edges at infinity
i.e. by imposing $L\sim N^{1/2}$. It is only by keeping $L$ fixed and sending $N$ to infinity that the truncated
model can be considered as gapped. It should be noted that the truncated model has eigenstates that are always lower
in energy than the magnetoroton branch hence its gap curve in Fig.(\ref{Gap}) is always lower than curve II.

We expect that the interactions between the edges become relevant when their spacing
is of order of the magnetic length. This spacing is $\Delta X\approx 2\pi N_{max}/L$ and the condition is thus $\Delta X=O(1)$.
At fixed filling factor this means $N|\log\lambda|^{1/2} =O(1)$. So when one increases $\lambda$ towards 1 we expect
a critical value $\lambda_{crit}=1-O(1/N^2)$ beyond which the physics involve interacting edges. This is not what we observe so far.
For accessible numbers of particles there is a value of $\lambda_{crit}\approx 0.8$ or $L_{crit}\approx 18\ell$
which is quite independent of the
number of particles. We thus expect that larger system sizes should show the critical value due to a crossing of levels 
to move towards $\lambda =1$.

%%%%%%%%%%%%%%%%%%%%%%%%%%%%%%%%%%%%%%%%%%%%%%%%%%%%%%%%%%%
\section{Edge states}
\label{edgecount}

Edge states have a counting which is simple in the disk geometry with a single chiral mode. They are generated
by acting on the Laughlin wavefunction with elementary power sums~:
\be
S_k =\sum_i z_i^k
\ee
The polynomials $S_k$ can be used to generate a basis of the symmetric polynomials in $N$ variables by taking all possible products.
For each partition of an integer $M$ written as $M=\sum_k km_k$ we construct the polynomial $\prod_k S_k^{m_k}$ which has
degree $M$. If this factor multiplies a wavefunction of fiducial degree $M_0$ then the state we obtain has total momentum $M_0+M$.
The number of such factors is the unconstrained partition number $p(M)$ (Sloane integer sequence A000041).
The Laughlin state is a compact droplet centered at the origin and the edge states generated by this process describe gapless
periphery deformations of the droplet. This set of states is in one-to-one correspondence with a free chiral boson.
The counting of edge modes pertaining to the $\nu=1$ droplet is exactly the same.
These modes have exactly zero-energy for the hard-core interaction and in the presence of a confining potential
they will give rise to a set of dispersive states corresponding to a boson with nonzero velocity.

In the cylinder geometry we have to consider the two boundaries. When the number of available orbitals is larger than the fiducial number to
``contain'' the given FQHE state then excitations may take place at the two edges. In the $\nu =1/3$ case, if we impose
$2N_{max}=3(N-1)$ as dictated by the shift quantum number of the FQHE state then there is no free space for edge motion
and the spectrum has an isolated ground state with a gap to bulk excitations. We
consider the situation with $N_{max}=m(N-1)/2+N_{\phi}/2$ with exactly $N_{\phi}$ additional flux quanta added to
the $\nu =1/m$ liquid. In the Landau gauge orbitals have a simple behavior in real space. They fully delocalized along the periodic
direction of the cylinder but exponentially localized in the orthogonal direction. The $\nu =1/m$ droplet can be pushed to the right
boundary of the cylinder and its wavefunction is given by~:
\be
\Psi^{(R)} = \prod_{i<j}(Z_i-Z_j)^m \times \left(\prod_i Z_i\right)^{N_{\phi}/2-m(N-1)/2}\, ,
\label{Rwave}
\ee
but it can be also pushed onto the left boundary~:
\be
\Psi^{(L)} = \prod_{i<j}(Z_i^{-1}-Z_j^{-1})^m \times \left(\prod_i Z_i\right)^{-N_{\phi}/2+m(N-1)/2}\, .
\label{Lwave}
\ee
These states are zero-energy ground states with opposite non-zero values of the total momentum.
They are related by action of the global shift operator ${\hat U}^{N_{\phi}}$ defined in Eq.(\ref{Uop}).
between the extremal values of the total momenta we find the globally translated ground states
${\hat U}^k$, $k=1,\dots N_{\phi}-1 $ but there are also all the states corresponding to edge excitations.
They can be generated by action of symmetric polynomials in the cylinder coordinates $Z_i$ admitting
positive as well as negative powers. If expanded in terms of the usual planar coordinates $z_i$
by the definition $Z_i=\exp (2\pi z_i /L)$ this is just a change of basis. The number of these states can be 
expressed through a bounded partition number ${\mathcal P}(N,M,D)$ where ${\mathcal P}$ is the 
number of partitions of $N$ into $M$ parts,
each part being at most $D$. Indeed zero energy states are given by~:
\be
(\sum_\sigma Z_{\sigma(1)}^{\alpha_1} \dots Z_{\sigma(N)}^{\alpha_N})\Psi^{(m)}\, ,
\label{zero}
\ee
where $\sigma$ are permutations and the powers $\alpha_i$ are bounded by $N_{\phi}/2$~:
$|\alpha_i|\leq N_{\phi}/2$. Hence their number is given by ${\mathcal P}(K+N(N_\phi/2+1),N,N_\phi+1)$.
The edge modes can also be counted from the root partitions of the wavefunctions. For example
if we consider the truncated Hamiltonians Eqs.(\ref{TF},\ref{TB}) then one obtains all zero-energy ground states by adding
any number of extra zeros in the root configuration. The number of ways to insert these extra unoccupied orbitals
is then simply given by the binomial coefficient $  \left(\begin{array}{c}  N+N_\phi    \\  N\end{array}\right)$.
This is simple only if we don't take into account the restriction of the fixed total momentum. That this number
of states is the same as the number obtained through counting symmetric polynomials is due to the following
combinatorial identity~:

\be
\left(\begin{array}{c}
   N+N_\phi    \\  N
\end{array}\right)
=\sum_{Q=0}^{Q=NN\phi}{\mathcal P}(Q+N,N,N_\phi+1)\, .
\label{combinat}
\ee

%%%%%%%%%%%%%%%%%%%%%%%%%%%%%%%%%%%%%%%%%%%%%%%%%%%%%%%%%%%
\section{Luttinger behavior in a parabolic well}
\label{Lut}

\begin{figure}
\includegraphics[width = 0.8\columnwidth]{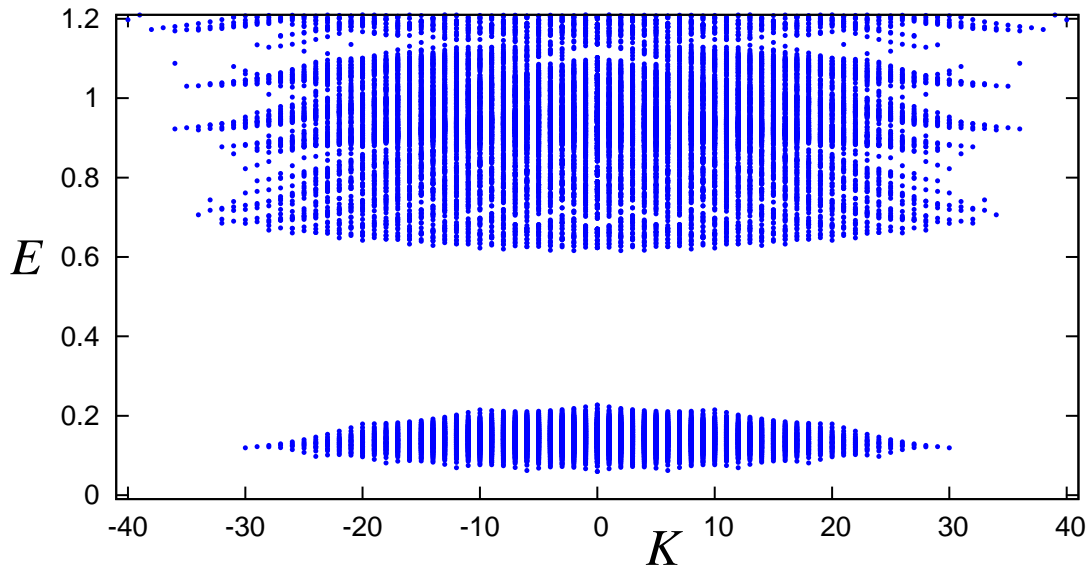}
\caption{Full spectrum of N=6 fermions on a cylinder with $\lambda =0.75$ for hard-core interactions as a function of the total momentum.
The confining potential has a strength $\beta=0.0008$ which ensures a clear separation between bulk and edge states.
Here there are 23 available orbitals while the Laughlin $\nu=1/3$ would need exactly 13 orbitals.}
\label{fullsp}
\end{figure}

\begin{figure}
\includegraphics[width = 0.8\columnwidth]{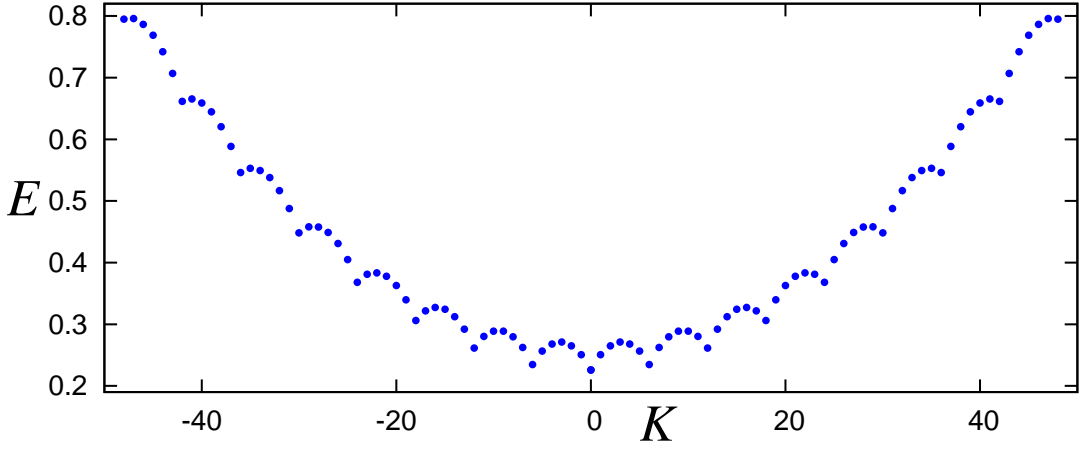}
\caption{The lowest energy state in a parabolic well as a function of the total momentum.
There are N=6 fermions in 36 orbitals.
The structure is typical of a finite-size Luttinger liquid. The states located at cusps
differ by a global translation of the occupation number pattern. They can be labeled
by the value of the total current $J=2nk_F$. The other states are internal excitations describable
by the phonons of the chiral Luttinger liquid if one stays close enough to the cusps.
Fitting the parabolic shape with Eq.(\ref{LutK}) leads to an estimate of the Luttinger parameter.}
\label{Arches}
\end{figure}

We now add a parabolic confining potential along the axis of the cylinder. In the Landau gauge
it is diagonal in the occupation number basis~:
\be
{\mathcal H}={\mathcal H}_{int} + \beta \sum_i i^2 {\hat n}_i
\label{Potential}
\ee
For small enough values of the coefficient $\beta$, the edge modes will remain well separated
from the bulk excitations but they will acquire dispersion. A typical spectrum is shown in
Fig.(\ref{fullsp}) where we have chosen a very small value of the trapping potential. The degeneracy of the edge
states that we discussed in the previous section is now lifted. When the edges are separated in space
by a large distance with respect to the magnetic length unit we expect that the two chiral modes
living on each extremity of the cylinder will be essentially decoupled. If we reduce this separation
by sending $L\rightarrow \infty, \lambda\rightarrow 1$ then the two edges will interact and form
a more complex Luttinger liquid. In the decoupled case, the total action is just the sum of two chiral
edge theories with the same Luttinger parameter $g$. Within the set of edge modes there are two
striking facts~: there is an overall low-energy parabolic envelope (see Fig.(\ref{Arches}))  and there are 
extremal states for $K=Np$
where $p$ is an integer.

%%%%%%%%%%%%%%%%%%%%%%%%%%%%%%%%%%%%%%%%%%%%%%%%%%

\begin{figure}
\includegraphics[width = 0.8\columnwidth]{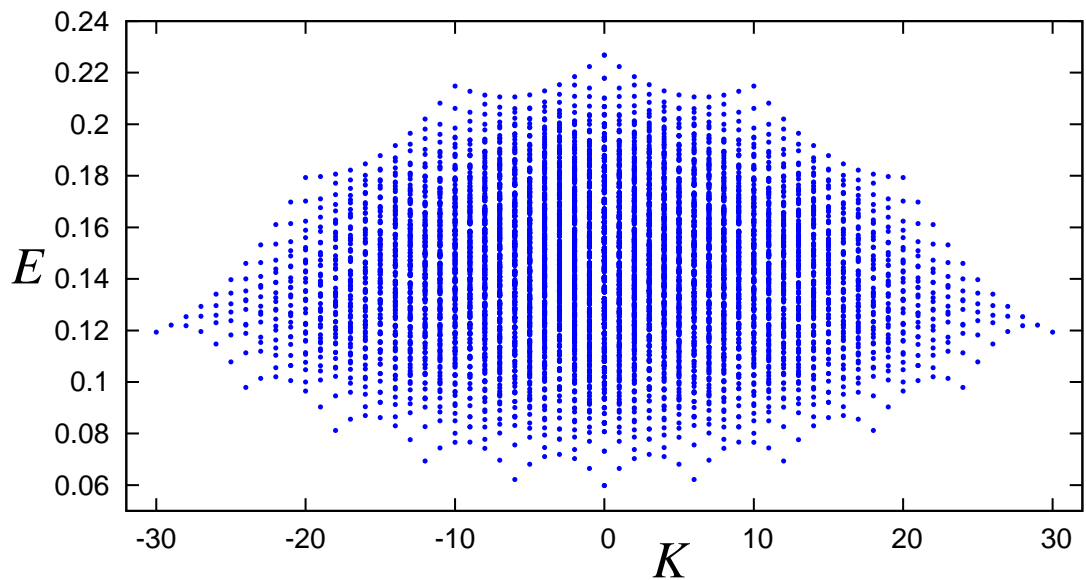}
\caption{A close-up of the previous picture. We focus only on the edge modes allowed by the extra 10 orbitals.
The states with extremal values of the total momentum correspond to the liquid droplet compressed against
the cylinder boundaries. The counting of edge modes matches the formulas given in section (\ref{edgecount}).}
\label{edgemodes}
\end{figure}

%%%%%%%%%%%%%%%%%%%%%%%%%%%%%%%%%%%%%%%%%%%%%%%%%%%%
These two features are exactly what we expect in a \textit{non-chiral} Luttinger liquid~\cite{Haldane81}.
In a finite-size system the non-chiral Luttinger liquid spectrum~\cite{Haldane81} is given by the following
effective Hamiltonian theory~:
\be
{\mathcal H}_L=v\sum_{q\neq 0} |q|\, b_q^\dag b_q + \frac{\pi g}{2L}vJ^2 ,
\label{LutK}
\ee
\be
K=k_F J +\sum_q q \, b_q^\dag b_q
\ee
where the operators $b_q^\dag b_q$ create bosonic modes with a velocity $v$ and the Fermi momentum
is given by $k_F=N/L$. The integer quantum number $J$ is even and counts the number of particle-hole excitations
across the Fermi sea while the boson operators keep track of the small-energy excitations on a given side
of the Fermi sea.
If we consider the edge spectrum of Fig.(\ref{edgemodes}) we can clarify the nature of the excitations by going to the
TT limit. The extremal states are now simple Slater determinants and we find that they differ exactly by a global translation
as can be done by acting with operator $\hat U$.

Its effect can thus be easily computed in the Tao-Thouless limit. 
The edge states span the zero-energy subspace in the absence of external potential.
We just have to add the effect of the parabolic potential Eq.(\ref{Potential})
onto this subspace. Since it is diagonal in occupation numbers its effects are obtained in an elementary
way. To obtain the lowest-energy one-phonon excitation one has to move a particle from the highest
occupied orbital in the root configuration to its neighboring empty site~:
\be
0\dots 001001001001\dots \rightarrow 0\dots 010001001001\dots
\ee
The one-phonon excitation energy is thus~:
\be
E_{1ph}= \beta [ (p_{max}+1)^2 - p_{max}^2 ]
\ee
where $p_{max}$ is the momentum of the highest occupied orbital $p_{max}=m(N-1)/2$.
So we have $E_{1ph}=\beta Nm$ and since this is $2\pi v/L$ in the Luttinger liquid theory
we obtain the value of $v=\beta Nm/(2\pi L)$.
Now the $J$-type excitations are given by an overall shift of the root configuration.
The energy is thus~:
\be
E_J = \beta \sum_i [ (p_i + \frac{J}{2})^2 - p_i^2 ] = \beta N \left( \frac{J}{2}\right)^2
\ee
where $p_i$ are all the momenta that appear in the root configuration.
With the expression of the Hamiltonian including the zero-mode contribution
we obtain the result given by Wen's theory $g=1/m$. In the TT limit there is no effect of the statistics
so this holds for Fermi as well as Bose cases.
So the simple TT limit has the correct energetics to reproduce the non-trivial exponent
of the FQHE edge theory. 

Away from this limit we resort to the following procedure to evaluate the Luttinger parameter~:
we take as the reference energy the $K=0$ state and fit the first arch with momenta
$K=0,\dots ,N$ by a polynomial of order two. 
The linear part in $K=0$ gives our estimate for $v$. Then 
the overall parabolic envelope gives an estimate of the zero-mode contribution
in Eq.(\ref{LutK}). Eliminating the phonon velocity gives an estimate of the Luttinger parameter.

\begin{figure}
\includegraphics[width = 0.8\columnwidth]{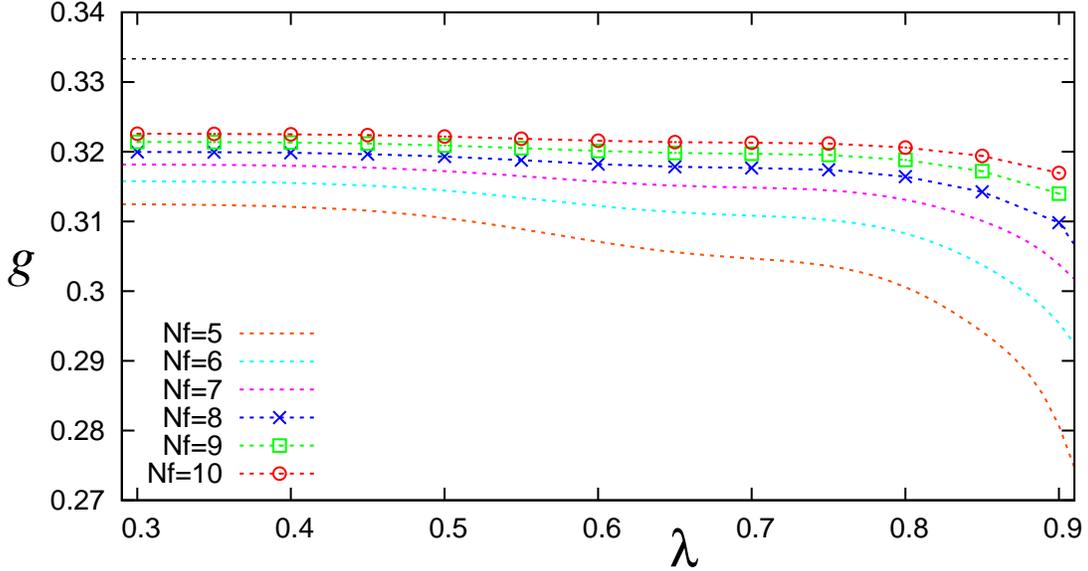}
\caption{Values of the Luttinger parameter as a function of the distance between edges on a cylinder
in the fermionic $\nu=1/3$ case. From Wen theory we expect the asymptotic value $g=1/3$.
Deviations are expected to occur when the edges are very close hence $\lambda\rightarrow 1$.}
\label{KFermi}
\end{figure}

\begin{figure}
\includegraphics[width = 0.8\columnwidth]{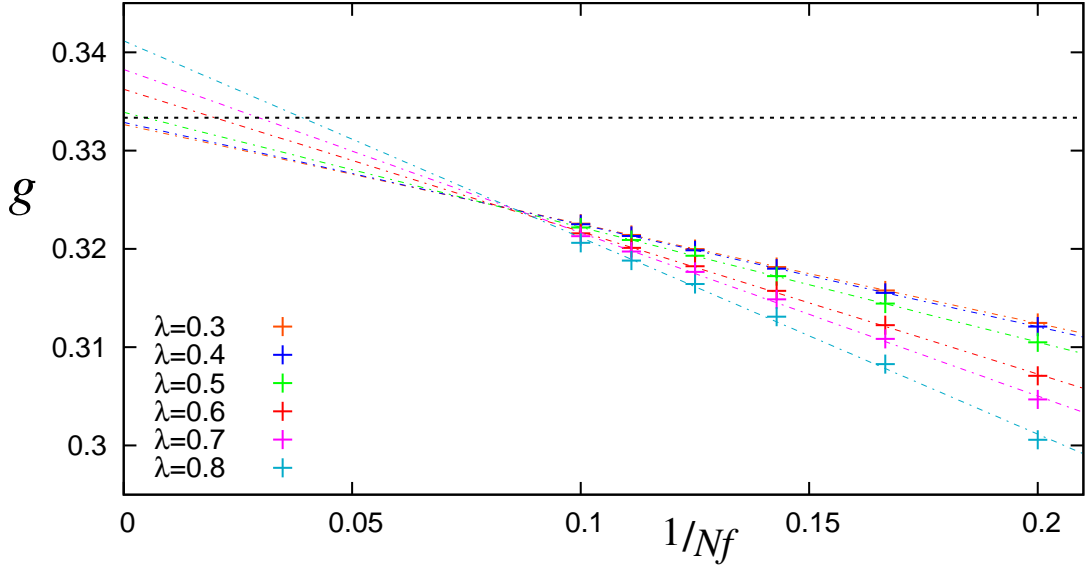}
\caption{Extrapolation of the Luttinger parameter to the thermodynamic limit in the Fermi $\nu=1/3$ case.
Values are plotted as a function of 1/N where N is the number of fermions.}
\label{LK}
\end{figure}

\begin{figure}
\includegraphics[width = 0.8\columnwidth]{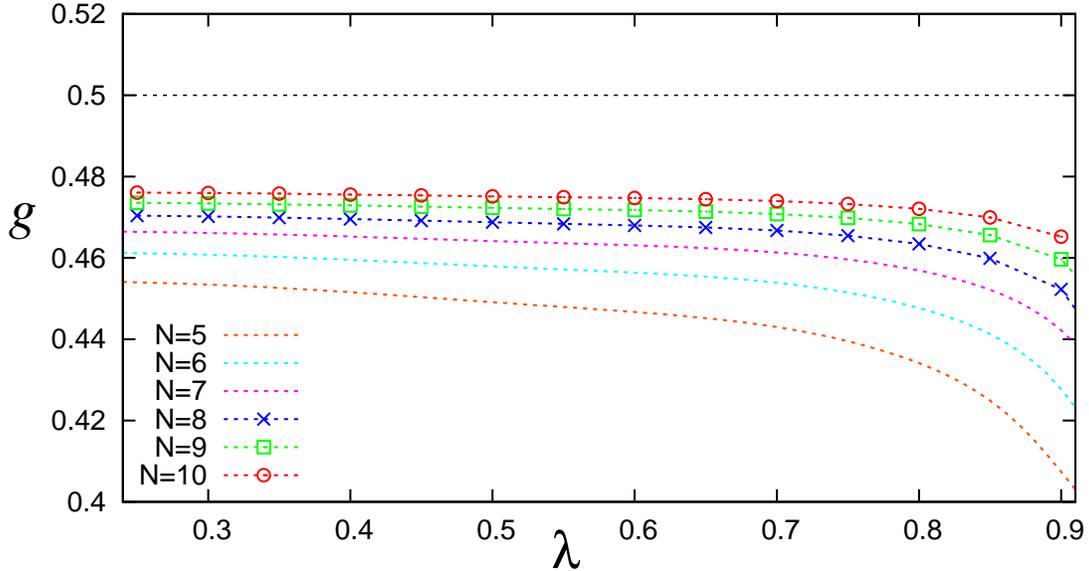}
\caption{The edge Luttinger exponent in the bosonic $\nu=1/2$ case following the same procedure as in the
Fermi case. There is good agreement with
the edge theory value $g=\nu$.}
\label{KBose}
\end{figure}

Our results are given in Fig.(\ref{KFermi}) as a function of the cylinder parameter $\lambda$ and the number of
fermions  between 5 and 10 at $\nu =1/3$ for the pure hard-core interaction.
We observe a wide regime with convergence towards the value $g=1/3$. This convergence can be
appreciated in Fig.(\ref{LK}) where the data at fixed $\lambda$ are plotted as a function of 1/N
and a tentative linear fit is consistent with the value 1/3. Similar results holds for bosons at $\nu=1/2$
with hard-core interactions~: see Fig.(\ref{KBose})

%%%%%%%%%%%%%%%%%%%%%%%%%%%%%%%%%%%%%%%%%%%%%%%%%%%

We have also performed a similar analysis for the truncated Hamiltonians Eqs.(\ref{TF},\ref{TB}) and found also that the edge exponent is
unchanged. This is to be expected since even in the TT limit it stays fixed at the chiral Luttinger value.
In the realm of the FQHE the effective theory leads to an edge electron propagator with an anomalous decay exponent
different from the Fermi-liquid case because the electron operator is a vertex operator when expressed as a function
of the Bose field. We will see that the TT state as well as truncated state do not have the correct spatial decay
of the electron propagator.

%%%%%%%%%%%%%%%%%%%%%%%%%%%%%%%%%%%%%%%%%%%%%%%%%%%%%%%%%%%
\section{The hoop limit}
\label{hoop}

In the graphs of the previous section there are deviations of the Luttinger parameter from
its FQHE value when $\lambda\rightarrow 1$. This is the ``hoop'' limit where all orbitals are collapsing
in space. Indeed if we stay at a fixed ratio of number of particles vs number of orbitals
sending $L\rightarrow\infty$ leads to vanishing edge separation. When considering the Hamiltonian in second-quantized
formulation we see that the interactions becomes very simple both for Bose and Fermi cases
after Fourier transformation.

\begin{figure}
\includegraphics[width = 0.8\columnwidth]{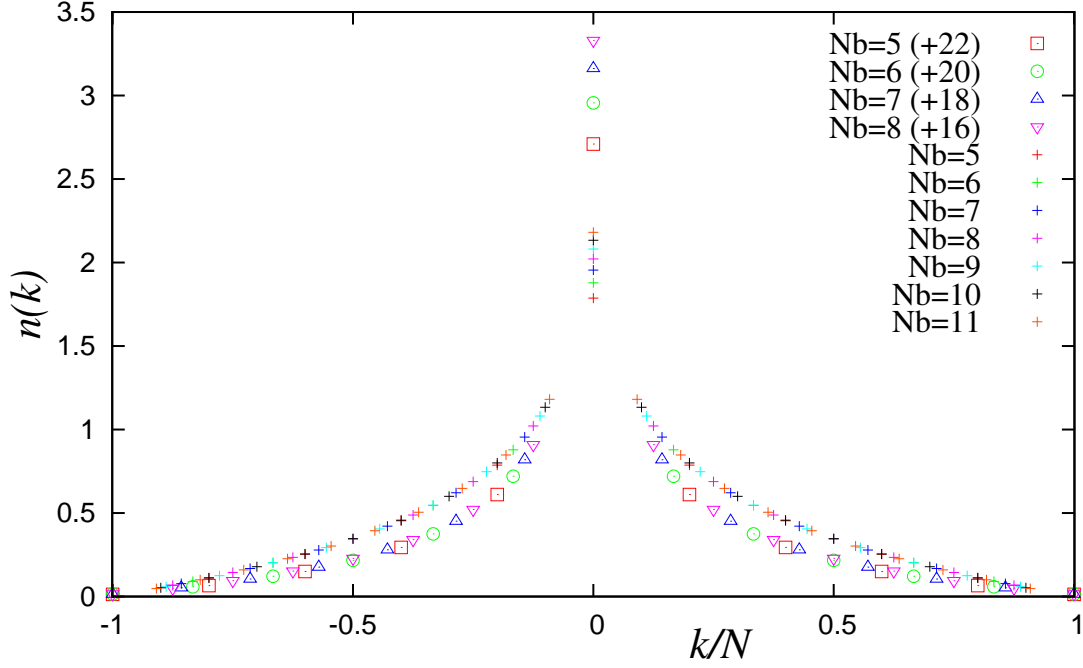}
\caption{The bosonic occupation number in the $L\rightarrow \infty$ limit. Without extra orbitals the limiting
behavior has a logarithmic singularity at $k=0$ the data are given for N=5 to 11 bosons. Extra orbitals are added
for N=5,6,7,8 and their number is given in parenthesis.
We observe a crossover to
a Tonks-Girardeau gas with a $k^{-1/2}$ singularity which is consistent with our limited data.}
\label{NkB}
\end{figure}

In the Bose case, it is proportional to the Hamiltonian of the Lieb-Liniger~\cite{Lieb63} one-dimensional Bose gas 
written in the plane wave basis~:
\be
{\mathcal{H}}_{LL}=-\sum_i \frac{\partial^2}{\partial x_i^2} + c\sum_{i<j}\delta (x_i-x_j),
\ee
provided there is no restriction on the available momenta. This restriction is fundamental because it leads to the special
Laughlin expression when we impose the relation $2N_{max}=2(N-1)$. If we add a large number of orbitals beyond
this relation, one should recover the physics of the Lieb-Liniger gas. Since we treat the case of a very small parabolic
potential in the sense that it should be small with respect to the strength of the interactions, it means that we are in the
so-called Tonks limit $c\rightarrow\infty$. There is thus fermionization~\cite{Girardeau60} and the Luttinger parameter approaches unity.
This is compatible with our numerical findings that the value obtained from finite-size effects increases when 
$\lambda\rightarrow1$ even though we cannot claim to be quantitative. There is a subtlety concerning the occupation number distribution
that should be noted. Even though the spectral properties are Fermi liquid-like the wavefunction is the absolute value
 of a Slater determinant of plane waves and hence it is not easy to compute the distribution $n(k)$. The exact available result~\cite{Lenard64}
shows that there is a zero-momentum singularity that replaces the Bose condensation forbidden in one dimension with
$n(k)\sim k^{-1/2}$ when $k\rightarrow 0$. This singularity is stronger than the logarithmic divergence of $n(k)$ when the $\nu=1/2$
Laughlin is exactly enclosed in its minimal number of orbitals. Our results are shown in Fig.(\ref{NkB}). They
are at least qualitatively consistent with the expected change of the singularity when adding orbitals.

\begin{figure}
\includegraphics[width = 0.8\columnwidth]{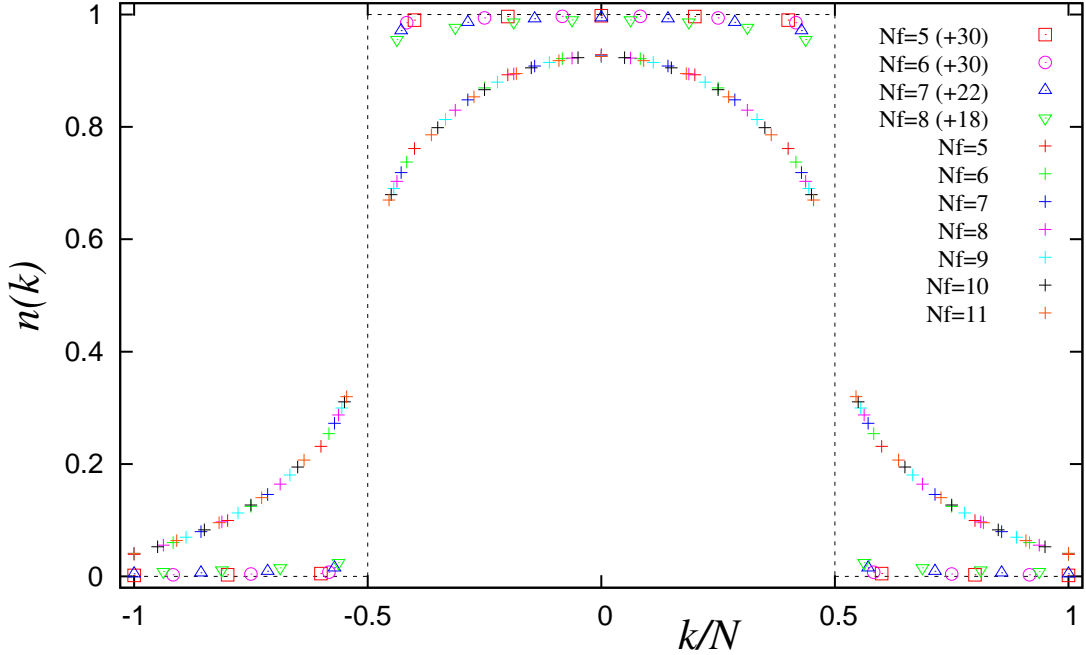}
\caption{The occupation number $n(k)=\langle c_k^\dagger c_k\rangle$ in the fermionic ground state 
at zero edge separation with and without extra orbitals. The crosses stands for the case without extra orbitals.
We recover the Calogero-Sutherland profile observed by Rezayi and Haldane. Note the prominent singularity
at $k=k_F$. Extra orbitals are added for N=5,6,7,8 and their number is given in parenthesis.
The singularity is washed out and replaced by a Fermi-liquid like discontinuity.}
\label{NkF}
\end{figure}

In the Fermi case, the Hamiltonian Eq.(\ref{HamF}) in the hoop limit is proportional to~:
\be
\mathcal{H}_{hoop}=-\int dx\, \Psi^\dagger \Delta \Psi \, +\, 
c_f\int dx\, \Psi^\dagger\nabla\Psi^\dagger \Psi \nabla \Psi ,
\label{ftg}
\ee
where we have introduced fermion operators $\Psi(x)$ for the coordinate along the edge i.e. along the hoop.
This is the simplest Pauli-allowed local one-dimensional interactions between spinless fermions
provided one uses in the definition of second-quantized operators the LLL basis functions.
So this may be called a fermionic Tonks-Girardeau gas~\cite{Girardeau03,Girardeau04}. But the definition of this last
model requires a space of functions which is not the same as ours.
When we release the constraint
on orbital number by adding extra available states we see that the smooth singularity at $k=k_F$ present
in the Laughlin state develops into a full discontinuity when there are many added states. 
This can be seen in our Fig.(\ref{NkF}).

The smooth singularity at $k_F$ observed from the Calogero wavefunction~\cite{RH94}
has an exponent equal to 2/3~: $n(k)\approx (k-k_F)^{2/3} $ as analyzed by Yang et al.~\cite{Yang96}.
It is not the chiral edge singularity which is found at $k=mk_F$~: $n(k)\approx (k-mk_F)^2 $. 
The discontinuity at $k_F$ that we observe is Fermi-liquid-like.

%%%%%%%%%%%%%%%%%%%%%%%%%%%%%%%%%%%%%%%%%%%%%%%%%%%%%%%%%%%
\section{Edge correlations}
\label{edgecorr}

In this section we compute several physical properties of the truncated states and discuss their difference
with the physics of the FQHE states.

\subsection{Momentum distribution of the truncated model}

We first compute the occupation numbers for the special truncated ground states introduced in section II.
We discuss the fermionic $\nu=1/3$ case in view of its interest for the physics of edge states. 
It is convenient to note that the ground state wavefunction Eq.(\ref{exact}) can be defined recursively~:
\be
|\Psi_N\rangle =|100\rangle \otimes |\Psi_{N-1}\rangle
- 3\lambda^4 |011000\rangle\otimes|\Psi_{N-2}\rangle
\ee
This recursion allows a simple computation of the occupation numbers.
Inside the bulk, i.e. when the site index $3p$, $N$, and $N-3p$ tend to infinity, the
asymptotic value of the momentum distribution
is then given by~: 
\begin{equation}
 n^{Bulk}(3p) = \frac{1}{\sqrt{1+36\lambda^8}}, \quad n^{Bulk}(3p+1) =
n^{Bulk}(3p+2) =
\frac{1}{2} \left(  {1-  \frac{1}{\sqrt{1+36\lambda^8}}} \right)
\end{equation}
The period-3 of the momentum distribution is similar to the Tao-Thouless
limit ($n(3p)=1,~n(3p+r)=0$). Contrary to the TT limit there is a value
of $\lambda=(2/9)^{1/8}$ for which the occupation number distribution is uniform.
Near the edges, i.e. when $N$ tends to infinity whereas $3p$ stays finite, the
momentum distribution has also a simple form~: 
\begin{equation}
 n^{Edges}(i) = n^{Bulk}(i) \left( {1- \left( 
\frac{1-\sqrt{1+36\lambda^8}}
{1+\sqrt{1+36\lambda^8}} \right)^{\lfloor{i/3} \rfloor + 1}}
\right)
\label{nedge}
\end{equation}
It is important to note that even in the thermodynamic limit this distribution
does not have the smooth vanishing property that we expect from
the chiral Luttinger liquid description $n(k)\sim (k-k_F)^2$.

\subsection{Density distribution and propagator of the truncated model}

We next focus on the density distribution 
$\rho(x,y)=<\Psi^\dagger(x,y)\Psi(x,y)>$ of the ground state of the truncated model. Since the total momentum
along the y-direction is a good quantum number, the density distribution can
be written as~: 
\begin{equation}
 \rho(x,y) = \frac{1}{L \sqrt{\pi}} \sum \limits_i e^{-(x-x_i)^2} ~
n(i),
\end{equation}
where $x_i= {2 \pi i}/{L} $. The density distribution is
homogeneous along the $y$-direction because of translation invariance,
and each orbital indexed by $i$ has a  spread along the  $x$-direction given by the Gaussian factor. In the
thermodynamic limit, the distance between
orbitals goes to zero, which leads to the homogeneous bulk density of  a liquid with filling factor $\nu =
1/3$~:
\begin{equation}
  \rho^{Bulk}(x,y) \equivalent_{L \rightarrow \infty} \frac{1}{3}\frac{1}{2 \pi}
\end{equation}
At finite $L$ there are density oscillations along the $x$-direction in the bulk density profile.
These oscillations are in fact present in all cases~: the TT state, the truncated state as well as the
Laughlin state. This crystal-like behavior disappears in the thermodynamic limit.

There is however a difference between these three states relative to the density profile at the edge.
In the TT state there is no bump at the edge for any system size.
There is a bump in the density at the edge for finite size in the truncated and Laughlin state.
This bump does not survive in the thermodynamic limit for the truncated model whereas it
does persist in the Laughlin case.

%%%%%%%%%%%%%%%%%%%%%%%%%%%%%%%%%%%%%%%%%%%%%%%%%%%%%%%%%%%
\subsection{Electron propagator}

The propagator of the electron along the edge of the system can also be computed in a similar way. 
If $x_e$ is the $x$ coordinate near the edge, the electron  propagator 
can be written~: 
\be
 \mathcal G (y) = <\Psi^{\dagger}(x_e,y)\Psi(x_e,0)> 
  = \frac{1}{L \sqrt{\pi}} \sum \limits_j e^{-(x_e-x_j)^2} e^{- i \frac{2 \pi}{L} j y} n(j) .
\ee

We now take the thermodynamic limit as explained in section \ref{geo}. We first send $N$ to infinity, which
 implies that the momentum distribution has to be taken near the edges Eq.(\ref{nedge}). We note that
the second term in equation  Eq.(\ref{nedge})
is exponentially decreasing with a characteristic length of order $1/L$. Its contribution to the propagator is then  negligible 
when $L \rightarrow \infty$, and only the first part $ n^{Bulk}$ contributes. One finds for $N \to \infty$ and $L\to \infty$~: 
\be
 \mathcal G (y)  \sim  \frac{1}{L \sqrt{\pi}} \sum \limits_{p=0}^{+\infty} 
e^{-(\frac{2 \pi }{L} 3 p)^2\texttt{}} e^{- i \frac{2 \pi}{L} 3 p y} 
\sim \frac{1}{12 \pi} e^{-\frac{y^2}{4 }}\, \mathrm{Erfc} \left( i\frac{y}{2 } \right),
\ee
where Erfc is the complementary error function. With the asymptotic expansion of the Erfc function, we obtain the behavior of the 
equal-time propagator of the truncated model at long distance : 
\begin{equation}
 \mathcal G (y) \equivalent_{|y|\to \infty} \frac{1}{y}
\end{equation}

So, we find that the equal-time propagator along the edges of the truncated model is identical in the thermodynamic limit to that of the 
Tao-Thouless model, and has the same asymptotic exponent as the propagator along free fermion edges. It does not display the characteristic 
exponent $\alpha =3$ of the Laughlin wavefunction. We note that more detailed investigations of the non-equal time correlations
would require the use of exact diagonalization techniques for dynamical properties~\cite{OG93}.

%%%%%%%%%%%%%%%%%%%%%%%%%%%%%%%%%%%%%%%%%%%%%%%%%%%%%%%%%%%
\section{Conclusions}
\label{conclude}

We have studied several features of the FQHE physics in the cylinder geometry by using exact diagonalizations as well as by
using exact results available on a truncated version of the Hamiltonian. The geometry we use is specially suited
to the study of edge states which can be generated easily by adding extra orbitals with respect to the fiducial number
given by the filling factor and the shift quantum number of the state under consideration. The use of hard-core interactions
allows us to cleanly separate  edge modes from bulk states. So this is restricted to FQHE fractions for which one can
construct a model Hamiltonian whose exact zero-energy ground state is the model state we want to study. This
includes Laughlin principal fractions $\nu=1/m$ for bosons and fermions. We have shown that the bulk gap can be obtained
at fixed flux as in the sphere geometry. However one has to fine-tune an extra parameter which is essentially
the equivalent of the aspect ratio of the torus geometry. This is what we have called $\lambda$.

By addition of orbitals we have shown that the standard counting of edge states is realized in the cylinder geometry with
now two counterpropagating edges at each end of the system. Since they combine to form a non-chiral Luttinger liquid
we have shown that the addition of an external potential can be used to measure the Luttinger exponent introduced
in Wen's theory. The same method can be applied in the TT limit analytically and the truncated Hamiltonian can also be
analyzed. In all cases we find the same edge exponent, which means that these models have correct energetics.
The edge combination takes place in real space~: it has been also observed in the entanglement spectrum~\cite{Lauchli2010}.

When the cylinder radius goes to infinity then the physics becomes one-dimensional and the problem is
a Tonks-Girardeau type model if we add a confining potential. In both Fermi and Bose cases there are 
signatures of a free fermion behavior.

However a key feature of FQHE liquids is the non-Fermi-liquid  behavior of the electron propagator at the edge.
The propagator exponent is known to be $\alpha =m$ for the Laughlin wavefunction, for $m=2$ and $m=3$. 
We have shown that the TT state as well as the truncated state have the Fermi-liquid exponent $\alpha =1$.
So the correspondence between the exponent in the spectral properties and in the algebraic decay of the correlations is broken
in these special cases.
This means that even if they capture some of the energetics of the FQHE these model states miss the delicate topological
correlations built in the FQHE state which are present in the Laughlin wavefunction.

%%%%%%%%%%%%%%%%%%%%%%%%%%%%%%%%%%%%%%%%%%%%%%%%%%%%%%%%%%%%%%%%%%%%%%%
\begin{acknowledgments}
We acknowledge discussions with N. Bray-Ali, G. Misguich, V. Pasquier, as well as  P. Roche.
\end{acknowledgments}

%%%%%%%%%%%%%%%%%%%%%%%%%%%%%%%%%%%%%%%%%%%%%%%%%%%%%%%%%%%%%%%%%%
%\bibliography{edges}
%%%%%%%%%%%%%%%%%%%%%%%%%%%%%%%%%%%%%%%%%%%%%%%%%%%%%%%%%%%%%%%%%%%%%%%

%%%%%%%%%%%%%%%%%%%%%%%%%%%%%%%%%%%%%%%%%%%%%%%%%%%%%%%%%%%%%%%%%%%%%%%
\end{document}